# Ocular Diseases Diagnosis in Fundus Images using a Deep Learning: Approaches, tools and Performance evaluation

Yaroub Elloumi [a,b,c], Mohamed Akil [a,*], Henda Boudegga [b]

[a]Gaspard Monge Computer Science Laboratory, ESIEE-Paris, University Paris-Est Marne-la-Vallée, France; [b]Medical Technology and Image Processing Laboratory, Faculty of medicine, University of Monastir, Tunisia; [c]ISITCom Hammam-Sousse, University of Sousse, Tunisia.

## ABSTRACT

Ocular pathology detection from fundus images presents an important challenge on health care. In fact, each pathology has different severity stages that may be deduced by verifying the existence of specific lesions. Each lesion is characterized by morphological features. Moreover, several lesions of different pathologies have similar features. We note that patient may be affected simultaneously by several pathologies. Consequently, the ocular pathology detection presents a multi-class classification with a complex resolution principle. Several detection methods of ocular pathologies from fundus images have been proposed. The methods based on deep learning are distinguished by higher performance detection, due to their capability to configure the network with respect to the detection objective.

This work proposes a survey of ocular pathology detection methods based on deep learning. First, we study the existing methods either for lesion segmentation or pathology classification. Afterwards, we extract the principle steps of processing and we analyze the proposed neural network structures. Subsequently, we identify the hardware and software environment required to employ the deep learning architecture. Thereafter, we investigate about the experimentation principles involved to evaluate the methods and the databases used either for training and testing phases. The detection performance ratios and execution times are also reported and discussed.

## 1. INTRODUCTION

Several ocular pathologies have a higher disease rate in the world. Their prevalence is moving to increase due to the population ageing and the unbalanced diet. Many of them lead to the blindness such as the Diabetic Retinopathy (DR), glaucoma, the Aged- macular degeneration (AMD), etc. Moreover, some late stages are irreversible chronic pathology, whatever may be the ocular therapy. Thus, the ophthalmologists invite patients to periodically screen with the aim of detect an eventual disease. In this context, several research works are interested in the automatic detection of ocular pathologies. Machine learning-based methods offered higher performance detection, especially those based on deep learning.

In fact, the fundus images illustrate several retinal components such as the blood vessel tree, the optic nerve head, the macula, etc. which have affine shapes and morphologies. The ocular pathologies lead either to reform retinal components or/and appearance of lesions. Those lesions differ in terms of size, shape, contrast, etc. Moreover, their always have similar characteristics than other retinal components or other pathological lesions. Therefore, the ocular diseases diagnosis seems to be difficult task, that requires taking into account several parameters, and hence the Deep Learning (DL) represents an adequate approach to resolve such problems.

This paper corresponds to an overview of DL-based methods for ocular pathology detection. These methods are always composed by a fundus image preprocessing step before DL processing. However, we distinguish a great diversity on the preprocessing in terms of principles and objectives. Moreover, several works employ the DL-architecture as the unique machine learning to achieve the target result. Other ones proceed to join several machine learning approaches. Therefore, we describe in the first section the preprocessing step of DL-based methods which is classified into two categories respectively for image enhancement and for data augmentation. Thereafter, we present the relation between the DL architecture and the input data management. The third section focuses on the description of the methods which are split

This work was supported by the PHC-UTIQUE 19G1408 Research program.

* Mohamed.akil@esiee.fr; phone +33 (0)6 73 53 46 89; fax 33(1).60.95.75.75; ligm.u-pem.fr/accueil/

## 2. PREPROCESSING

**2.1 Preprocessing for fundus image enhancement**

In most instances, methods based on DL proceed to apply preprocessing before the training step. The preprocessing provides resized fundus images to be conformed to the expected input of DL network. We distinguish reduced resolutions such as 180*180 [16], 224*224 [8], 227*227 [5], 231*231 [2], middle resolutions such as 416*416 [9], 512*512 [6], and higher resolutions like 1200*1800 [4].

The preprocessing objective varied with respect to the methods. Some methods propose preprocessing to enhance fundus image quality. The work described on [4] proceeds to apply histogram equalization to balance the intensity difference across images. Wan et al. [17] aims to enhance fundus image by applying Nonlocal Means Denoising (NLMD). Grassmann et al. [6] method consists of normalizing the color balance as well as local illumination of each fundus image by using a Gaussian filtering to subtract the local average color. In [11], the original images are converted from RGB to Hue Saturation Intensity color space, and then denoised using median filter. Thereafter, CLAHE method is applied to enhance contrast. In [15], the fundus image is converted from RGB to LUV color space. The L channel is extracted to remove varying local contrast and enhance illumination. Then, the image is converted back to RGB color space.

Other methods lead to upgrade lesions or ROI (Region of Interest Detection) that correspond to the aimed pathology, in the objective of assisting training and hence increasing the performance detection. The method proposed in [4] aims to detect Retinal Microaneurysm (MAs). In this context, the method consists of applying mean filter to the green channel. Then, top-hat and Gaussian filter are applied to distinguish MAs from blood vessels. After that, a threshold is employed to extract MA candidates. The work proposed by [10] consists of applying Gaussian filter and morphological operator in order to detect red lesion candidates from fundus image, before applying DL. The method proposed in [11] aims to detect the optic disk when detecting hard exudates. Therefore, it proceeds to apply Laplacian of Gaussian (LoG) filter to the green channel, and then a circular convolution mask with radius similar to those of the optic disc. Thereafter, a mask is applied to provide the OD location.

**2.2 Preprocessing for data augmentation**

The data augmentation consists of modifying fundus image in order to rise the data set size and to increase robustness for the DL model. Within this objective, fundus images rotation is always employed several times with randomly angle [17, 6, 8]. Another data augmentation corresponds to flipping or mirroring fundus images. The orientations are chosen randomly in the method proposed in [6], while [8] employs horizontal and vertical flips and [16] flips image in left, downwards and left then downwards orientation.

The image cropping is also employed which consists of modifying shape size in the fundus image [6, 9, 8, 17]. Another data augmentation consists of adjusting shapes, where fundus images are randomly adjusted between 0% and 15% in [6], 10% of pixels are removed in [9], and zoomed in [17, 8]. In addition, the method proposed by [9] proceeds to noising, blurring, modifying lightness and brightness of fundus images.

## 3. DL-BASED METHODS VS DATA INPUT MANAGEMENT

We deduce an important gap between sizes of input images using in deep learning architectures. The image input sizes are in relation with the method objective, which can be partitioned into two categories.

The first one corresponds to the methods aiming to detect pathologies or pathological severity. These methods proceed to only resize input images with respect to the deep learning architectures. The work of [2] proposed to detect the AMD (Aged-related Macular Degeneration) and classify the image as early, intermediate or advanced stages. The images are cropped into square and then resized to 231x231. The work proposed by [6] consists to detect the AMD Degeneration and their severities throw deep learning method where images are resized to 512*512. In [16], the authors aim to detect AMD using DCNN (Deep Convolutional neural network) where images input is rescaled to 180x180. In [Chai et al. 2018], the authors propose to detect glaucoma and segment glaucoma related areas from retinal image throw « multi-branch neural network model ». The images are cropped into squares and then resized to 192*192. The work described by [5] proposes to detect cataract from retinal image where images are preprocessed to enhance the quality and resized to 227x227. We deduce that images are always narrowed, which leads to significant lost on image features. Moreover, we deduce an important difference between input image sizes, and even for the same pathology detection. However, even works offer higher performance detection, no work has been proposed to quantify the impact of feature lost, due to the image resizing, on the performance detection.

The second category corresponds to methods aiming to segment ocular lesions. Those methods proceed to squaring the input image into patches. The work of [1] proposes to segment HE (Hard Exudates) using neural network architecture. The images are split into patches of 32x32 to be classified into two class hard exudates and background. The work of [7] proposes a method for automatic exudates detection using multiple deep learning architectures.

The retinal images are cut into patches of size 25x25 with three color channels which were labeled in two groups: (i) Exudates and (ii) Non-exudates. The work described by [11] proposes to combine the output of optic disc and vessel detection with the output of the convolutional neural network in order to detect exudates. The images used for training are divided into sub-images with 65*65 pixels. In [9], the authors propose to localize the region of interest to analyses glaucoma using deep convolution neural network.

At first, the images datasets are resized to 416*416 and then cut into patches of size 13x13. Shan et al. [13] propose to detect microaneurysms throw deep learning network where images are cut into patches of dimension 25x25 and then injected into auto-encoder architecture.

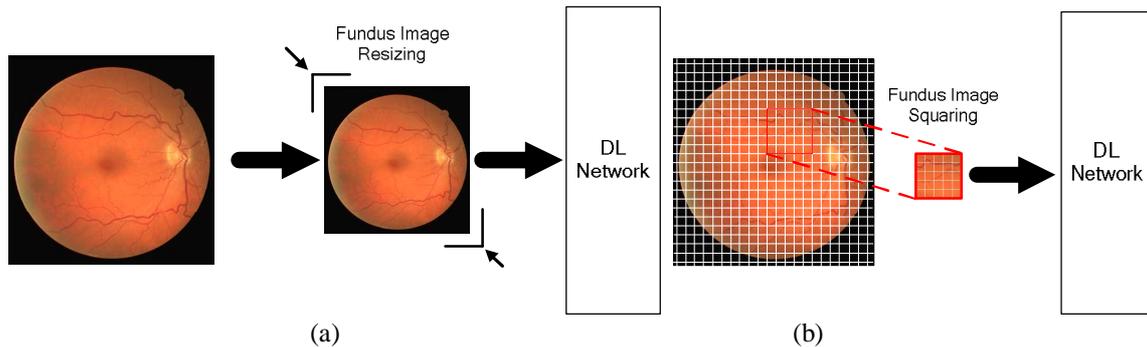

(a) (b)
Figure 1. Data input for DL network on ocular pathology detection methods : (a) fundus image resizing ; (b) Fundus image squaring

The squaring allows increasing dataset sizes used for the training which always leads to enhance detection performance. Benzamin et al. [1] proceed to extract 5000 patches from each image, where 2500 are HE patches and 2500 are background patches. The work proposed in [13] consists at extracting 70 patches from each image. In fact, several studies described lesion morphologies and offered size range of each lesion whatever the pathology severity is. However, the patch sizes are not explicitly chosen in terms of lesion sizes. Besides, we deduce a main difference between patch sizes, even for detecting the same lesions.

## 4. MAIN PROCESSING OF DL-BASED METHODS

**4.1 Fully DL-Based methods**

The work described in [1] splits each image in 32*32 patches that are introduced to a CNN network in order to classify the centered pixel as hard exudates. The work described in [2] leads to detect AMD using a DCNN that has been pre-trained on ImageNet. Each image is cropped and resized and then classified as early stages/intermediate or intermediate/advanced stages. A method proposing ROI (Region of Interest Detection) coordinates including class probabilities of glaucoma is proposed in [9], which are based on CNN comprising 24 layers of convolution, batch normalization and max pooling surfaces.

The glaucoma can be directly detected through a method based on 18-layers CNN, proposed in [12]. Tan et al. [16] was proposed a fourteen-layer deep CNN model to diagnose AMD at an early stage. The diabetic retinopathy lesions exudates, hemorrhages, micro aneurysms are able to segment through a 10-layer CNN [15]. The work described on [17] proceeds to apply several CNNs to the same data set in order to classify a fundus image into 5 classes of DR, whose are respectively AlexNet, VggNet, GoogleNet and ResNet. Then, it evaluates their detection performances, without proposing an optimal model. The work described on [8] proposes a method based on DL network as a 5-class classification of DR.

**4.2 Partially DL-based methods**

The work described in [3] leads to detect glaucoma using several CNN models partitioned into three branches. The first one branch consists of detecting the glaucoma directly by injecting fundus image through deep learning models. The second branch uses Faster-RCNN to obtain optic disc region. For the third branch, Faster CNN is used to segment disc area, cup area and Parapapillary Atrophy (PPA) area. The work described in [4] identifies microaneurysms candidate through an automatic image-to-text mapping. Then, a multi-scale convolutional neural network (MS-CNN) is processed to filter the false positives candidate. The work described in [6] proposes a method of 9 classes of AMD disease, which employs multiple CNN models independently. Then, a random forest algorithm is trained based on results provided by CNNs. Another work leads to automatically detects the exudates was proposed in [7]. It employs several deep learning methods which are respectively the CNNs, pre-trained Residual Networks (ResNet-50) and Discriminative Restricted Boltzmann Machines. Moreover, the softmax layer in the ResNet-50 is replaced with three different classifiers: OPF, SVM, and k-N other work described in [10] leads to segment red lesions. After enhancing lesion shapes, a CNN is trained to provide a feature vector. Then, A Random Forest (RF) classifier is trained using the provided feature vectors to avoid false positive candidate lesions.

## 5. PERFORMANCE OF DL-BASED METHODS FOR OCULAR DISEASES

We report in this section recent works that interested in ocular pathology diseases which we propose to split them into two categories that are respectively pathology detection and lesion segmentation. For the first category, some works aim to detect the disease in dependently form different lesions that brings, which are summarized in table.1. Other works proceed to prove the pathology disease by detecting one of their lesions. We notice that several works aim to define the severity of an ocular pathology, such as indicated in [5]. For the second category, the works consist of segmenting lesions caused by pathologies, where are shown in table.2. We distinguish that several works are focused on only one lesion whereas other ones lead to detect several lesions using the same DL model.

Both category works are validated either using a public fundus image database and local databases where pathological image are equitably partitioned with respect to the severity grades. We notice that DL models are evaluated using datasets containing fundus images with different pathologies in order to evaluate the robustness of the method, such as the work described in [3]. The large majority of methods propose a unique DL-architecture. Moreover, they proceed to evaluate performance based on different dataset. Consequently, it seems difficult to compare DL models based on their performance. We notice that few works offer a comparison of performance detection between several DL architectures, such as the one proposed by [17].

Table 1. Works of pathology/lesion detection

| Work | Pathology | DL model | DL architecture | Fundus image Data base | Performance detection |
|---|---|---|---|---|---|
| [2] | DMLA | Deep CNN | 19 layers + L-SVM | AREDS | • Acc:92-95%<br>• SP:89.8-95.6%<br>• SE: 90.9-96.4%<br>• PPV : 88.3-92.3%<br>• NPV:93.8-97.3% |
| [6] | DMLA | Alexnet GoogleNet VGG inception-v3 ResNet ResNet-V2 | {Alexnet, GoogleNet, VGG, inception-v3, ResNet, Resnet-V2} + Random forest | AREDS | • Acc: 57.7% - 85.7%<br>• K : 55.47% - 92.14% |
| [16] | DMLA | CNN | 7 convolution layers+ 4 max-pooling layers+ 3 fully-connected layers | Local dataset: 402 normal retinal images, 583 retinal images with intermediate AMD, or GA; 125 retinal images with evidence of wet AMD. | Blindfold:<br>• Acc: 91.17%<br>• Sen: 92.66%<br>• Spec: 88.56%<br><br>Ten-fold:<br>• Acc: 95.45%<br>• Sen: 96.43%<br>• Spec: 93.75% |
| [12] | Glaucoma | CNN | 4 Convolutional layer+ 4 Batch normalization layer+4 Rectified Linear | Local dataset:<br>• 589 normales images | • Acc: 98.13%<br>• SE: 98.00% |

| | | | Unit (ReLu)+ 4 max pooling+Fully connected layer+Soft-max layer | • 837 glaucoma images | • SP: 98.30%<br>• PPV: 98.79% |
|---|---|---|---|---|---|
| [5] | Cataract | CNN | Convolution+relu+pool+noarmalization+{SVM/softmax} | Local dataset: 7851 fundus images : 4671normal, 2176 mild, 622 medium, 382 severe | **softmax**:<br>4-class : ACC=90.82%<br>2-class : ACC=94.07%<br><br>**SVM**:<br>4-class : ACC=84.7%<br>2-class :ACC=89.83% |
| [8] | Diabetic retinopathy | VGG-D | Convolution+max pooling+full connected+dropout +sigmoid | EyePACKS | • Acc:81.7%<br>• SP :50.5%<br>• SE :89.5% |
| [17] | Diabetic retinopathy | AlexNet VggNet GoogleNet ResNet | • AlexNet:5 convolution layers+3 pooling layer+3 full connected layers+softmax<br>• VggNet: 5 convolution layers+5 max pooling layer+2 dropout+ 3 full connected layers+softmax<br>• GoogleNet: 22 layers deep CNN + inception layers | Kaggle | **AlexNet** :<br>• SP : 94.07%<br>• SE : 81.27%<br>• AUC : 93.42%<br>• ACC: 89.75%<br><br>**VggNet-s**:<br>• SP : 97.43%<br>• SE : 86.47%<br>• AUC : 97.86%<br>• ACC:95.68%<br><br>**GoogleNet**:<br>• SP : 93.45%<br>• SE : 77.66%<br>• AUC : 92.72%<br>• ACC:93.36%<br><br>**ResNet**:<br>• SP : 95.56%<br>• SE : 88.78%<br>• AUC : 93.65%<br>• ACC: 90.40% |
| [7] | Hard exudates (Diabetic retinopathy) | CNN ResNet-50+ classifier DRBMs | • CNN: convolution+max pooling+normalization+full connected<br>• ResNet+{OPF, SVM, k-NN} | DIARETDB1 e-Ophtha | **DIARETDB1**<br>• Acc:98.2%<br>• SE:99%<br>• SP:96%<br><br>**e-Ophtha**:<br>• Acc:97.6%<br>• SE:98%<br>• SP:95% |
| [13] | Micro-aneurysms (Diabetic retinopathy) | Stacked sparse auto encoder | 2 layer stacked sparse autoencoder+ softmax | DiaretDB | • Presision:91.57%<br>• SP: 91.60%<br>• F-measure: 91.34%<br>• ACC:91.38%<br>• AUC:96.20% |

Table 2. Works of pathological lesion segmentation

| work | Pathology | DL model | DL architecture | Fundus image Data base | Performance detection |
|---|---|---|---|---|---|
| [15] | Exsudates Hemorrhages Micro-aneurysms(Diabetic retinopathy) | CNN | 3 convolution layers+ 3 max pooling+ 3 full connected layers+ softmax | Cleopatra | **Exudates** :<br>• SE : 87.58%<br>• SP:98.73%<br>**Haemorrhages**:<br>• SE : 62.57 %<br>• SP : 98. 93%<br>**Microaneurysms**:<br>• SE: 46.06 %<br>• SP: 97.99%<br>• |
| [1] | Hard exudates (Diabetic retinopathy) | CNN | 8 convolution layers+3 full connected + batch normalization + relu+ max pooling | IDRiD | • SE: 98.29%<br>• SP: 41.35%.<br>• ACC:98.6%. |
| [10] | Red lesions (Diabetic retinopathy) | CNN | 4 convolutional layers+ 1 fully connected layer+max pooling + dropout + avgpool | DIARETDB1<br>e-ophtha<br>MESSIDOR | **MESSIDOR**:<br>• AUC : 89.32%<br>• SE : 91.09% |
| [11] | Hard exudates (Diabetic retinopathy) | CNN | 4 convolution layers+ maxpooling+ 2 full connected layers | DRiDB | • Se: 78%<br>• PPV: 78%<br>• F-score: 78% |
| [3] | Glaucoma | • Faster RCNN FCN<br>• CNN | 5 convolution layers+max pooling+Dropout | Local dataset :2554 retinal fundus images :<br>* 1023 glaucomatous images:255 cataract, 173 diabetic retinopathy, 427 myopia, 142 macular degeneration<br>* 1531 normal images:334 cataract, 188 diabetic retinopathy, 365 myopia, 137 macular degeneration | • Acc :91.51%<br>• SP :90.90%<br>• SE :92.33% |
| [9] | Glaucoma | CNN | 24 layers:<br>• Convolution<br>• Batch Normalization<br>• Max Pooling | MESSIDOR<br>Kaggle<br>DRIVE<br>STARE | • **MESSIDOR**:Acc :99.0%<br>• **Kaggle**: Acc : 98.78%<br>• **DRIVE**: Acc : 99.41%<br>• **STARE**: Acc : 98.37% |

## 6. HARDWARE AND SOFTWARE ENVIRONMENTS

Recent works DL-based methods for ocular diseases are always implemented using TensorFlow library, either with or without Keras API, Caffe framework or Matlab. The three solutions have software tools to automatically perform the training on GPU architectures in order to benefit from parallel processing performance and hierarchical memory with the aim of decreasing the training execution time.

The Matlab tool allows implementing all steps required by the DL-methods both for preprocessing and training. However, even TensorFlow and Caffe insure data augmentation and squaring processing, they are limited to perform preprocessing to enhance input image such as applying filters, morphologic operators, etc. In this context, several works proceed to join the OpenCV library that represents an Open Computer Vision widely written in C++ [18, 19].

Table 3. Hardware and software environment of DL-method for ocular pathology diseases

| Work | Software environnement | Hardware environnement |
|---|---|---|
| [3] | Keras API for TensorFlow library | • Intel Xeon(R), 2.00 GHz E5-2620 CPU,<br>• NVIDIA GeForceGTX 1080 GPU |
| [4] | Caffe framework<br>Python , C++ | • Intel Xeon E5-2630 v4 @ 2.20GHz CPU<br>• NVIDIA Tesla K80 GPU |
| [5] | Caffe framework | K80 GPU in server |
| [7] | Caffe framework | NVIDIA Geforce GTX 1070 GPU |
| [8] | OpenCV<br>Keras API for TensorFlow library<br>CUDA and CuDNN libraries | Intel Core-i7 5820K, 64GB RAM<br>GeForce GTX 770 GPU |
| [9] | Keras API for TensorFlow library | --- |
| [11] | Caffe framework | Tesla K20C graphics card |
| [12] | MATLAB | Intel Xeon CPU E3-1225, 3.3GHz, 16GB RAM<br>GPU |
| [14] | MATLAB | CPU: 2.3 GHz, 4Gb RAM |
| [1] | Tensorflow library | --- |
| [15] | MATLAB | Intel Xeon 2.20 GHz (E5-2650 v4), 512GB RAM |
| [16] | Keras API for TensorFlow library<br>Theano backend | Intel Xeon 2.20 GHz (E5-2650 v4) , 512 GB RAM |
| [17] | CUDA code for training | GPU for training CNN |

## 7. CONCLUSION

In this paper, we presented an overview of DL-based methods for ocular pathology detection. We started by describing the preprocessing step where tasks are classified as enhancement and data augmentation processing. Then, we noticed how the fundus image is explored in order to be processed introduced the DL-architectures either by resizing and squaring on different patches. Thereafter, we partitioned the methods into fully DL-based methods and partial DL-based ones and we related their performances of the ocular disease detection and the hardware and software environments.

We deduce a great divergence between methods the processing, the architecture, the input data management and the performance evaluation; even several methods aim the same objective. Moreover, the majority of DL-based methods are interested to a unique ocular pathology. However, the clinical context requires detecting several eventual diseases in the same screening, which correspond to a real challenge.